\documentclass[english]{nature}
\usepackage[T1]{fontenc}
\usepackage{xcolor}
\usepackage{graphicx}
\usepackage{dcolumn}
\usepackage{bm}
\usepackage{hyperref}
\usepackage{amsmath}
\usepackage{amssymb}
\makeatletter

\usepackage{graphicx}
\makeatletter
\let\saved@includegraphics\includegraphics
\AtBeginDocument{\let\includegraphics\saved@includegraphics}
\renewenvironment*{figure}{\@float{figure}}{\end@float}
\makeatother

\begin{document}

\title{ Macroscopic Matter Wave Quantum tunnelling}

\author{Khemendra Shukla$^1$, Po-Sung Chen$^1$, Jun-Ren Chen$^1$, Yu-Hsuan Chang$^1$ \& Yi-Wei Liu$^{1,2*}$}
\maketitle

\begin{affiliations}
\item Department of Physics, National Tsing Hua University, Hsinchu 30013, Taiwan.
\item Center for Quantum Technology, National Tsing Hua University, Hsinchu, 30013, Taiwan.

\end{affiliations} 

\begin{abstract}
Quantum tunnelling is a phenomenon of non-equilibrium quantum dynamics and its detailed process is largely unexplored. We report the experimental observation of macroscopic quantum tunnelling of Bose-Einstein condensate in a hybrid trap. By exerting a non-adiabatic kick to excite a collective rotation mode of the trapped condensate, a periodic pulse train, which remains as condensate, is then out-coupled by quantum tunnelling. This non-equilibrium dynamics is analogue to tunnelling ionization. The imaged tunnelling process shows the splitting of matter wave packet by the potential barrier. The controversial ``tunnelling time" question is found inadequate, from the point of view of wave propagation. The realized matter wave pulse train can also be a passive pulsed atom laser for atom interferometer applications.
\end{abstract}

\section*{INTRODUCTION}
Quantum tunnelling, as a non-equilibrium process, is one of the most interesting problem in modern quantum physics. It has been investigated in microscopic systems such as the alpha decay of nucleus\cite{Holstein:1996il}, in quantum cosmology\cite{PhysRevD.15.2929}, tunnelling in Josephson's junction\cite{JOSEPHSON:1962vk}, and in biology\cite{AWSCHALOM:1992dl,Collini:2010fy}. Recently, the development of ultrafast laser allows the attosecond angular streaking and attoclock techniques to investigate the temporal momentum distributions of electron, based on the tunnelling ionization\cite{Eckle:2008eu,Pfeiffer:2011kl,Walt:2017hn,Kubel:2019ik,Sainadh:2019jw,Wu:2019fu}. However, quantum tunnelling has no counterpart in classical physics and its process remains mysterious. Particularly, the controversial tunnelling time problem is still unresolved\cite{Sainadh:2019jw,Torlina:2015dl,Ni:2016hq,Sokolovski:2018gx}. Thanks to the realization of Bose-Einstein condensate (BEC), the macroscopic quantum tunnelling (MQT) is now experimentally feasible. The interaction-assisted tunnelling of single and multi-particle has been studied experimentally\cite{Anderson:1998uf,Serwane:2011cv} and theoretically\cite{Lode:2012kr}. The tunnelling efficiency of BEC in an accelerating\cite{Lignier2007} or shaking\cite{Tayebirad:2010} periodic potential well has also been studied in momentum space. Such experiments have provided an unique approach to perform many-body physics experiments\cite{Bloch:2008ch} in an isolated environment, using controllable potential and inter-atomic interactions\cite{Potnis:2017dv}. Furthermore, to study the time evolution of the tunnelling process, the condensate, degenerate as a single macroscopic wave function and directly observable in spatial and temporal scales, is an excellent platform to study such non-equilibrium dynamics\cite{Lode:2012kr,Alcala:2017jj}. The laser cooled atomic BEC with tunability of the relevant parameters, such as the barrier height, the kinetic energy and the number of atoms in the tunnelling process, is advantageous for experimental realization.

 Our experiment is analogue to the strong field tunnelling ionization experiment, where the condensate acts as a bound electron wave packet. The potential formed by the optical dipole trap (ODT), the magnetic trap and gravity acts as the Coulomb potential of nucleus distorted by strong light pulse in the attosecond streaking technique. Using the macroscopic matter wave packet of BEC, the quantum tunnelling ``ionization" then can be studied in details. The distribution of tunnelling atoms through the trapping potential is a chronology of a collectively excited condensate inside the trap, as temporal imaging of valence electron motion\cite{Pfeiffer:2011kl,Walt:2017hn,Kubel:2019ik}.
 
 The tunnelling dynamics to open space has been demonstrated experimentally only with optical soliton\cite{Barak:2008bk}. In our report, the matter wave packet was used for the first time to exhibit the tunnelling dynamics to open space and to be observed in real space. On the other hand, as demonstrated in our report, tunnelling output atoms remained as condensates and were converted to quasi-one-dimensional matter wave pulses as a ``passive" pulsed atom laser, which has no active output coupling mechanism and can be utilized for atom interferometer applications.  

\section*{RESULTS}
\subsection{Tunnelling Matter Wave Pluses}

\begin{figure}
\centering
\includegraphics[width=1 \linewidth]{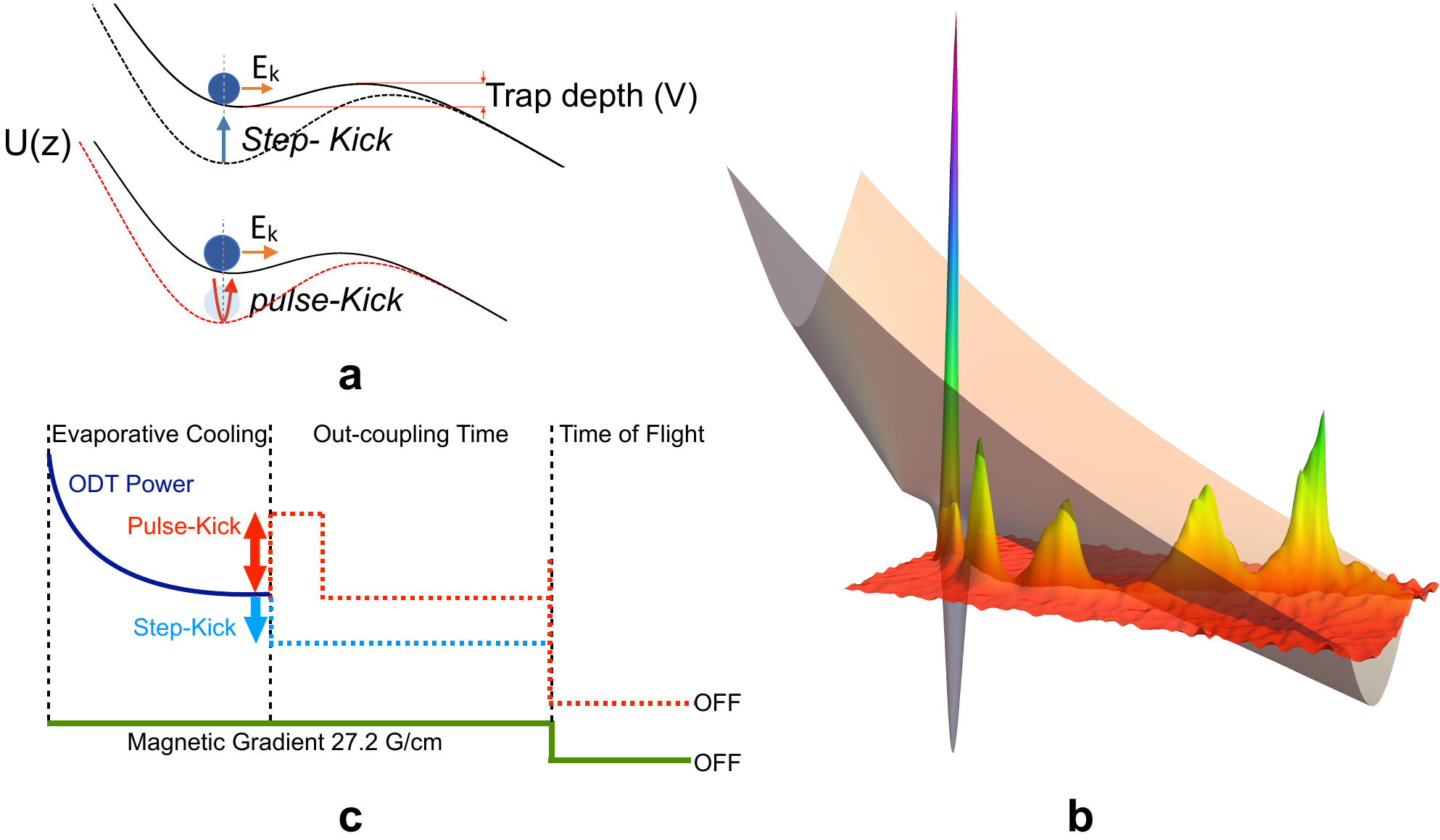}
\caption{\textbf{Tunnelling Potential and Time Sequence.} \textbf{a} The schematic representation of the potential well in z-direction for the non-adiabatic kicks: step-kick and pulse-kick. The change in the trap potential is done by varying the optical dipole trap (ODT) power. \textbf{b} The illustration for quantum tunnelling through the trapping potential (the grey shadow surface). The out-coupled matter wave pulses (the false color peaks) decay exponentially as described by the quantum tunnelling theory. \textbf{c} The operation time sequence is illustrated. The red dash line is the time-varying ODT power for the pulse-kick and the blue dash line is that for the step-kick. These kicks are implemented after creating Bose-Einstein condensate. The out-coupling time is the duration in which atoms are allowed to escape from the ODT.}
\label{fig:timeseq}
\end{figure}

The BEC was produced in a hybrid trap with a crossed ODT configuration, accompanied by a weak magnetic quadrupole field. Due to gravity, the potential barrier height becomes lowest at the bottom of the trap (see Fig.~\ref{fig:timeseq}a and~\ref{fig:timeseq}b). The field gradient of the magnetic quadrupole on the X-Y plane provides a transverse confinement, as shown in Fig.~\ref{fig:timeseq}b. The BEC was then perturbed by varying the trap depth of ODT potential. The protocols of the time-varying potential include: an adiabatic ramping and two types of kick (step-kick and pulse-kick), which are non-adiabatic changes in the potential well, as shown in Fig.~\ref{fig:timeseq}a and~\ref{fig:timeseq}c (see Method). 

\begin{figure}
\centering
    \includegraphics[width=1\linewidth]{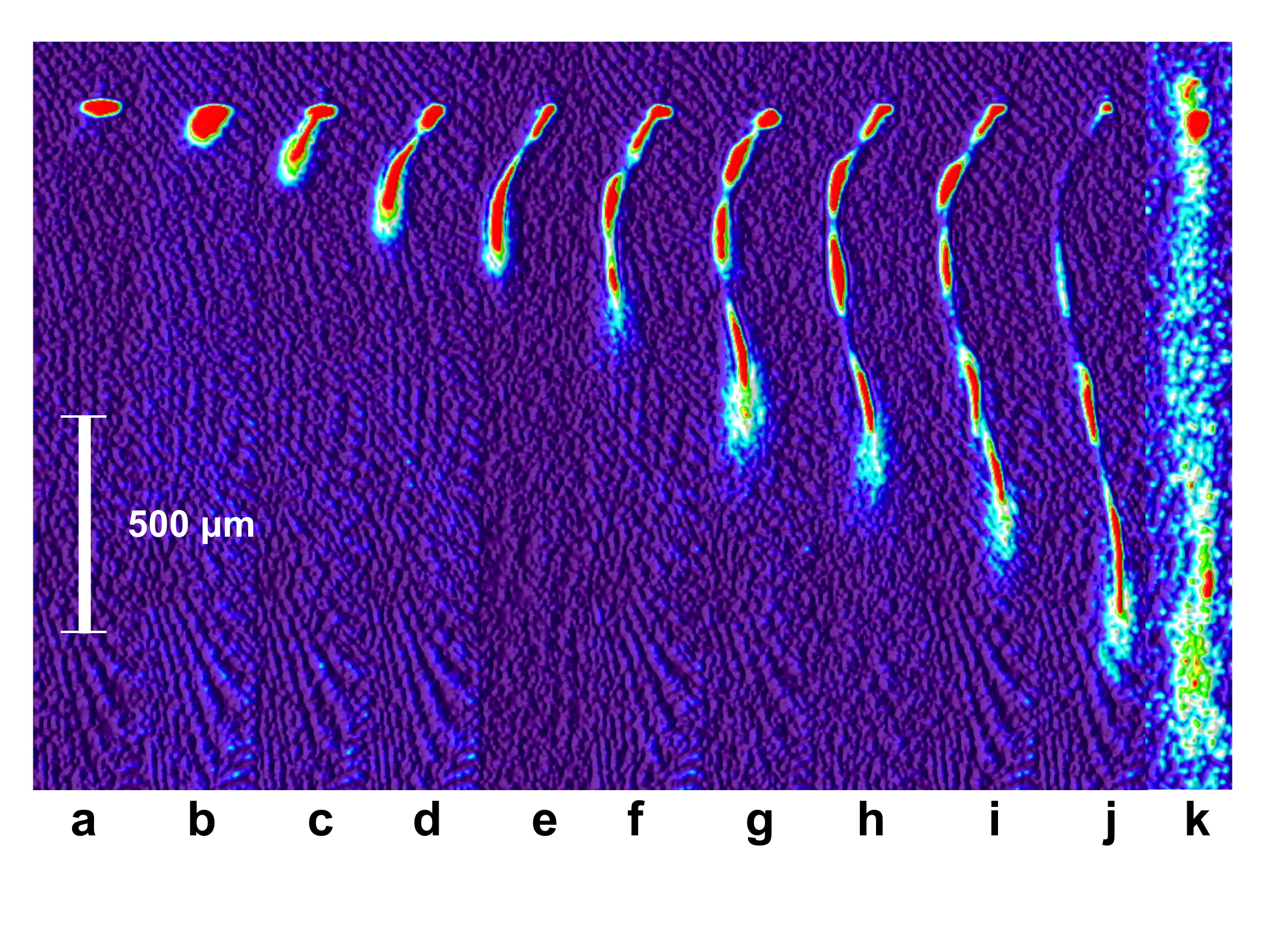}
    \caption{\textbf{Macroscopic Quantum Tunnelling Matter Wave Pulses.} \textbf{a} The absorption image of the Bose-Einstein condensate (BEC) created in the hybrid trap. \textbf{b}, \textbf{c}, \textbf{d}, \textbf{e}, \textbf{f}, \textbf{g}, \textbf{h}, \textbf{i} and \textbf{j} are the absorption images taken during the periodic tunnelling by applying a step-kick.   The atoms started to eject from the BEC reservoir \textbf{b} at 10 ms and \textbf{c} 20 ms after the kick. \textbf{d} The first complete  pulse with peak flux of $6.34\times10^6$ atom/s out-coupled at 30 ms after the kick. \textbf{e}, \textbf{f}, \textbf{g}, \textbf{h}, \textbf{i} and \textbf{j} are periodic pulses out-coupled at 40 ms, 50 ms, 60 ms, 70 ms, 80 ms and 90 ms after the kick, respectively. \textbf{k} After creating the BEC, the optical dipole trap power was reduced adiabatically and no kick was applied. It shows that the BEC was still held in the original position, and a continuous thermal atomic beam diffused out in both upward and downward directions.}   
\label{fig:growth}
\end{figure}

By exerting a step-kick, a series of periodic matter wave pulses was generated by quantum tunnelling as in Fig.~\ref{fig:growth}a  to~\ref{fig:growth}j, which shows a typical time evolution of the BEC dynamics. Using a step-kick, 4-5 pulses were ejected from the BEC reservoir in 90 ms as shown in Fig.~\ref{fig:growth}i and ~\ref{fig:growth}j. In cases with a larger condensate, the output pulsing can last for 200 ms to generate more than 10 matter wave pulses. The first pulse came out in $\sim$10 ms after exertion of kick (see Fig.~\ref{fig:growth}b). A complete pulse was ejected from the BEC reservoir in $\sim{30}$ ms with peak flux $6.34\times10^{6}$ atom/s as shown in Fig.~\ref{fig:growth}d. The output pulses are with a pulse duration of 20~ms defined by full-width-half-maximum, and a repetition rate of $\sim$50~Hz. The duty cycle is 50$\%$, which implies a symmetrical internal oscillating motion of the BEC reservoir. Considering the X-Y confinement by the magnetic quadrupole trap, the sinusoidal trajectory of the pulses also suggests an accompanied initial transverse momentum. Our experimental results agree qualitatively with the previous numerical simulation in several aspects, such as periodicity and pulse shape\cite{Salasnich:2002dr}.

On the other hand, when an adiabatic ramping was employed in comparison with a kick, a thermal atomic beam was produced. The ODT power was ramped down adiabatically in 2 sec to the same final trap depth as the step-kick. In such a case, neither additional kinetic energy, nor collective mode excitation was given to the condensate reservoir. A straight continuous diffusing leakage from the BEC reservoir was observed as shown in Fig.~\ref{fig:growth}k. It clearly showed that the final trap strength of ODT was still capable of holding the reservoir against the effective gravitational force (see METHOD). The diffusing leakage goes even above the position of BEC reservoir, and shows as a thermal atomic cloud, although this thermal cloud is still confined by the transverse trapping force. Therefore, we conclude that these are thermalized atoms energized by the ODT light assisted heating, and the condensate can tunnel through the trap only by applying a kick, even with such a low final potential barrier.

\begin{figure}
		\centering
		\includegraphics[width=1\linewidth]{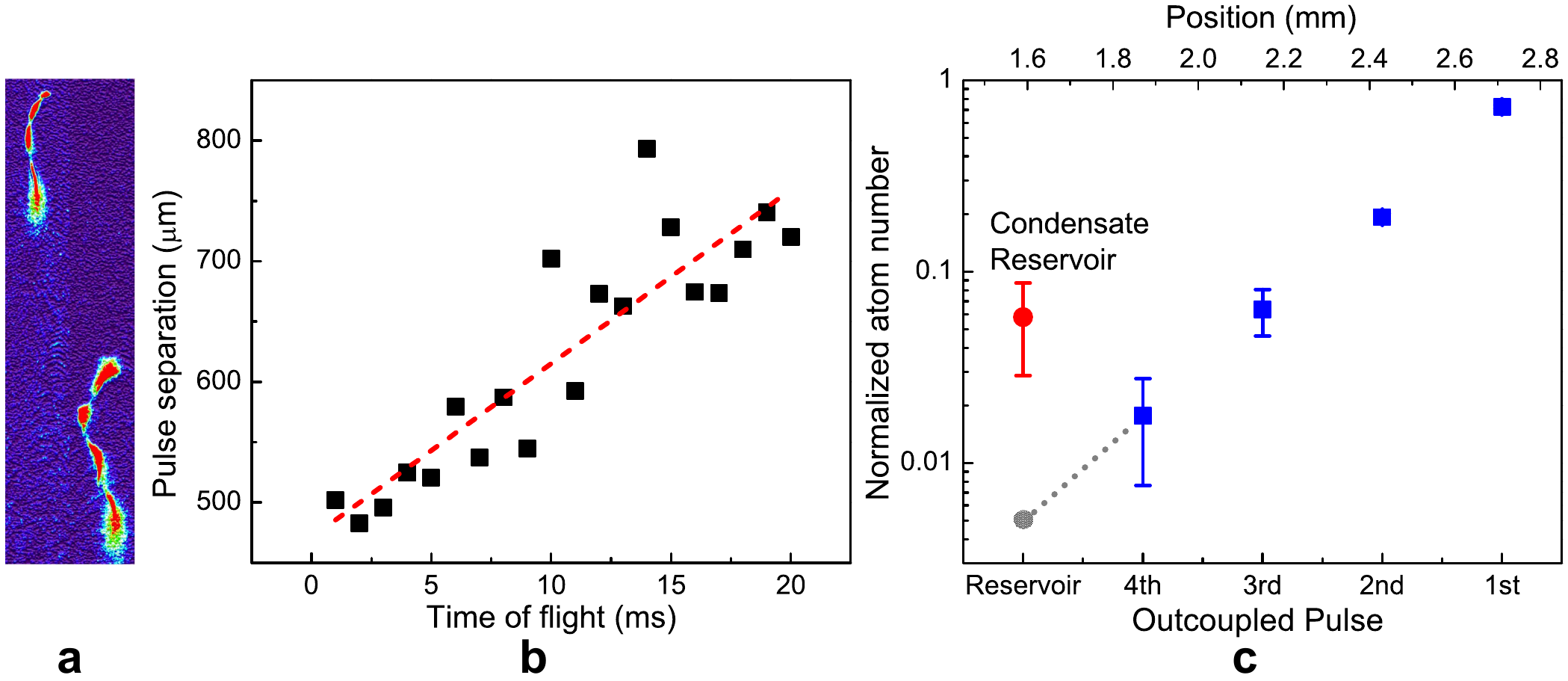}
		\caption{\textbf{Periodicity and Exponential Decay of Tunnelling Output Atoms.} \textbf{a} The comparison between the 1~ms and 20~ms time of flight (TOF) images of the matter wave pulses after 60 ms out-coupling time. Both the reservoir and the tunnelling output pulses remain as quantum degeneracy. No thermal expansion was observed. \textbf{b} The pulse separations versus the time of TOF. Each data point was obtained from single measurement. \textbf{c} The decay of the output pulses. The atom numbers of each pulse are normalized to the total ejected atom numbers. The bottom axis represents order of the output pulses and the top axis denotes the vertical position of pulses. The geometric progression decay, Eq.~(\ref{eq:coupling efficiency}), exhibits as a linear regression in the logarithmic scale. To compare with the measured residual atom in the condensate reservoir, the grey-dotted point is the extrapolated projection from the output pulses using Eq.~(\ref{eq:coupling efficiency}). The unexpected large residual in the reservoir indicates that $10\%$ of the entire condensate will eventually not tunnel out. The error bar was calculated from three measurements for each data point. }
		\label{fig:decay}
\end{figure}

To characterise the output pulses, the time of flight (TOF) measurement was performed for the three pulses generated in 60 ms of out-coupling time. The absorption images were taken after 20 ms free fall under gravity only, without the magnetic field, as shown in Fig.~\ref{fig:decay}a. There were certain amount of thermal atoms observed in the very beginning of the first output pulse. The rest of the sub-sequential pulses shows no thermal expansion that indicates them preserved as condensate. Fig.~\ref{fig:decay}b is a plot of pulse separation expansion with the free fall time. Its linearity implies that the pulses ejected from the BEC reservoir are periodic, and the pulse rate shows no slowing down.

In quantum tunnelling, the tunnelling efficiency $\eta$, the fraction of tunnelled atoms, only depends on the potential barrier $U(\textbf{r})$ and the kinetic energy of the matter wave $E_k$. Using the Wentzel-Kramers-Brillouin (WKB) approximation, it can be expressed as\cite{griffiths:quantum}:
\begin{eqnarray}
\eta=\exp{\left(-2\int_{\bm{r_1}}^{\bm{r_2}}\sqrt{\frac{2m}{\hbar^2}(U(\textbf{r})-E_k)}d\textbf{r}\right)}
\label{eq:tunnelling coefficient}
\end{eqnarray}
where $\bm{r_1}$ and $\bm{r_2}$ are the classical turning points on the both sides of the barrier. With a constant tunnelling efficiency $\eta$, the number of tunnelled atoms decays exponentially. It is an important characteristic of the quantum tunnelling. In our pulsed BEC experiment, the number of atoms in the $i^{th}$ output pulse is related to the tunnelling efficiency and given by:
 \begin{eqnarray}
 N_i=N_{0}\eta(1-\eta)^{i-1}
 \label{eq:coupling efficiency}
\end{eqnarray}
where $N_i$ is the number of atoms in $i^{th}$ pulse, $N_{0}$ is total number of the condensate. The exponential decay then exhibits as a discrete geometric progression. The Fig. \ref{fig:decay}c shows that the output pulses decay as described by Eq.(\ref{eq:coupling efficiency}), i.e., a linear line in logarithmic scale. Thus, the tunnelling efficiency $\eta$ is a constant, and we conclude that the kinetic energy of the pulses $E_k$, remains constant during the course of the 80 ms out-coupling time. The kinetic energy of atoms did not dissipate and the oscillation was not damped. Meanwhile, using Eq.(\ref{eq:coupling efficiency}), the expected total number of atoms remaining in the BEC reservoir after 4 pulse output is extrapolated and shown as the grey point connected by a dashed line in Fig. \ref{fig:decay}. However, the observed number of atoms that remain in the BEC reservoir (the red point) was  more than the extrapolated atom number (the grey point). It implies that 10\% of the entire BEC was not excited, or may transfer from the rotational mode to the vibrational mode at a later time, as described in the theory\cite{Chen:2005hl}.

\subsection{Tunnelling Efficiency}
\begin{figure}
\centering
		\includegraphics[width=1\linewidth]{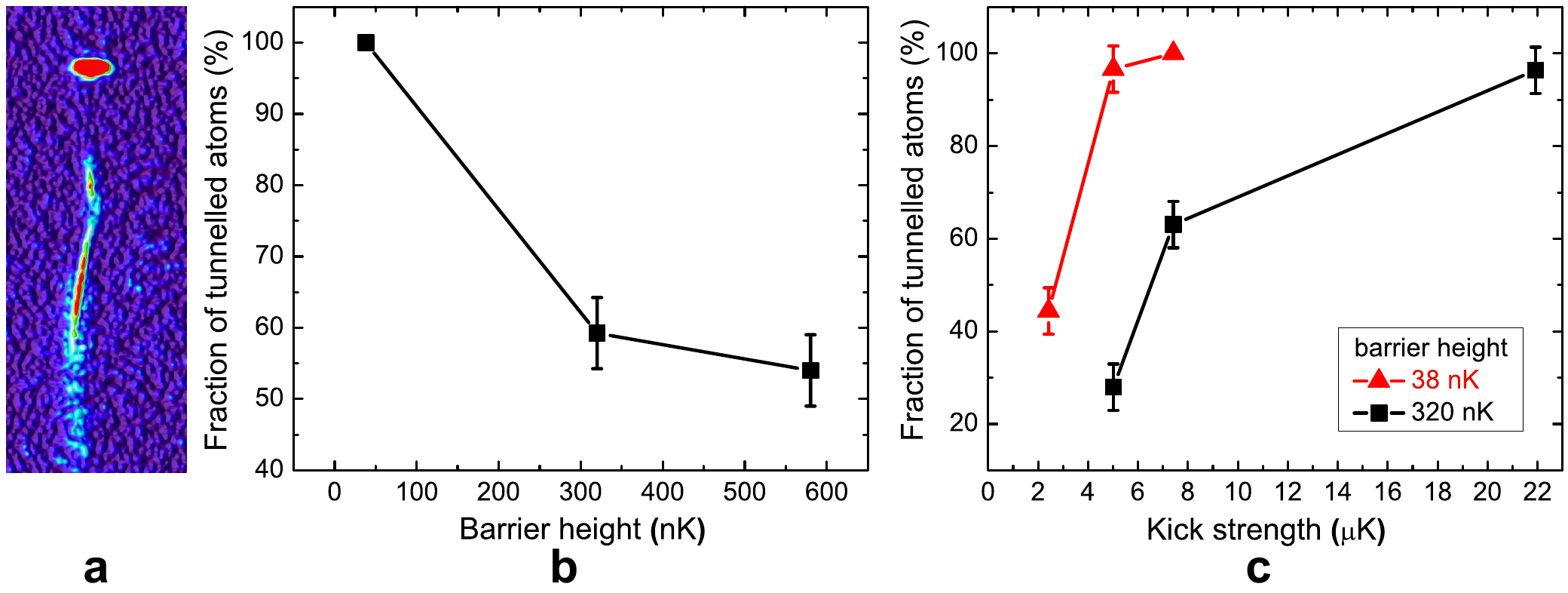}
		\caption{ \textbf{Tunnelling Efficiency.} The tunnelling efficiency (the fraction of tunnelled atoms) depends on barrier height and kick strength. Atoms are ejected with different efficiencies by applying a pulse-kick with various final barrier heights and kick strengths. The kick strength corresponds to the 2~ms sudden increment of the trap depth, and the barrier heights are the trap depth before and after the pulse kick. \textbf{a} A typical output pulse produced by a pulse-kick. \textbf{b} By varying the barrier height with a constant kick strength of 7.42 $\mu$K, the fraction of tunnelled atoms is plotted. At lower barrier height, large number of atoms can be ejected. \textbf{c} The kick strength was varied with the barrier heights of 38~nK and 320~nK. The error associated to a data point is calculated according to system reproducibility from three measurements.}
		\label{fig:kick}
\end{figure}

The pulse-kicks were applied to control the tunnelling efficiency $\eta$ ( Eq.~\ref{eq:coupling efficiency}). It allows us to set various final barrier heights $U(\textbf{r})$, and to tune the kinetic energy $E_k$ by varying the pulse height, the kick strength. Fig.~\ref{fig:kick}a is a typical absorption image of output pulse generated by a pulse kick. The tunnelling efficiency decays exponentially as the barrier height increases with a fixed kick strength, as shown in Fig.~\ref{fig:kick}b. The kick strength, expressed as the increment of the trap depth, is related to the kinetic energy $E_k$ given to the the condensate. The kick strength v.s. the tunnelling efficiency is shown in Fig.~\ref{fig:kick}c, where the final potential barrier height was kept as a constant.  While the pulse-kick gives an excitation energy higher than the barrier height, the output reach to 100$\%$. It shows an exponential saturation that qualitatively agrees with Eq.~(\ref{eq:tunnelling coefficient}). 

\subsection{Internal Dynamics}
\begin{figure}
		\centering
		\includegraphics[width=1\linewidth]{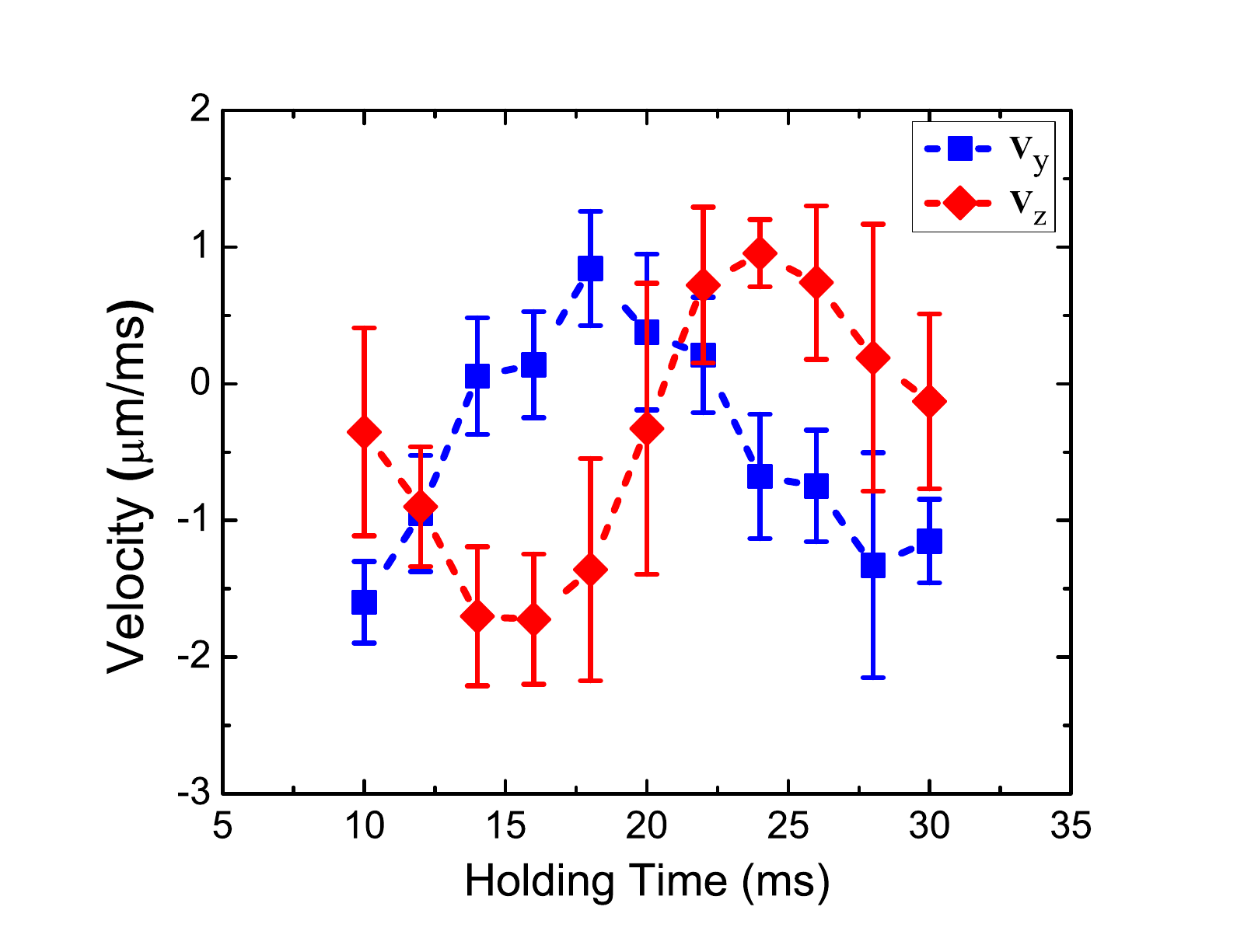}
		\caption{\textbf{Internal Dynamics of the Excited Condensate.} The internal dynamics of Bose-Einstein condensate by execution of the step-kick. The horizontal velocity $V_y$ and vertical velocity $V_z$ are the initial velocities of the condensate in the trap caused by kick. The velocities are extracted from the 10~ms TOF measurement of the atoms with various holding time after giving a step-kick, but without generating matter wave pulse. The error bar is calculated from three measurements for each data point.}
		\label{fig:dynamics}
\end{figure}

Although the trajectory of the output pulse in Fig.~\ref{fig:growth}j has mapped out the internal dynamics of the condensate reservoir, the conventional TOF method was utilized for further confirmation. The momentum oscillation inside the ODT, the excited collective mode, was mapped as the CM (center of mass) spatial oscillation after TOF. The result is as in Fig.~\ref{fig:dynamics}, which shows sinusoidal oscillations for both the velocities in y and z directions ($V_y $ and $V_z$), but with a $\pi/2$ relative phase shift.  The time period of oscillation is 20 ms which is in good agreement with the trap frequency. The amplitude of the velocity oscillation is $V_{z,y}=1.5~\rm{\mu{m}/ms}$ and the kinetic energy per atom is calculated to be 23.5 nK which is lower than the trap depth (38 nk). It is worthy to note that the created $\rm{^{87}Rb}$ BEC is with a sound velocity of 0.60 $\rm{\mu{m}/ms}$, and a chemical potential of 3.8~nK.

In a three-dimensional trap, the pure CM rotation mode, which is the lowest collective excitation mode with the rotation frequency $\omega=\omega_0$ and the angular momentum $l=1$\cite{Stringari:1996vi,Dalfovo1999}, evidently, is the most dominating mode by applying a weak perturbation. Using the TOF measurement, the observed trajectory and the distribution of atom number confirmed that the kick excites a collective rotational mode in the condensate with an energy lower than the potential barrier. The output matter wave pulse from the reservoir are due to quantum tunnelling. In this process, the output pulses and the reservoir itself remain in the phase of quantum degeneracy.

\subsection{Tunnelling Process}
The dynamics of BEC can be well described by the time-dependent Gross-Pitaevski equation (GPE) which is a mean-field approximation. While the non-linear interaction term in GPE,  $a_s|\psi(\bm r,\,t)|^2\psi(\bm r,\,t)$, becomes sufficiently small and negligible in comparison with the kinetic and potential energy terms, it becomes a standard time-dependent Schr\"{o}dinger equation. The BEC, which has a length scale of `$\rm{\mu m}$' and an evolution time scale of `ms', is an excellent object to study quantum tunneling. Unlike a single particle wavefunction that can only be fully explored by statistical measurement, the distribution of the BEC that exhibits the probability density by its number density can be imaged using a CCD camera by a single image. To resolve the controversial ``tunnelling time'' problem\cite{HAUGE:1989tm}, we hence performed quantum tunnelling using BEC, which is considered as a giant wave function with slow motion dynamics.

\begin{figure}
	\centering
	\includegraphics[width=0.9\linewidth]{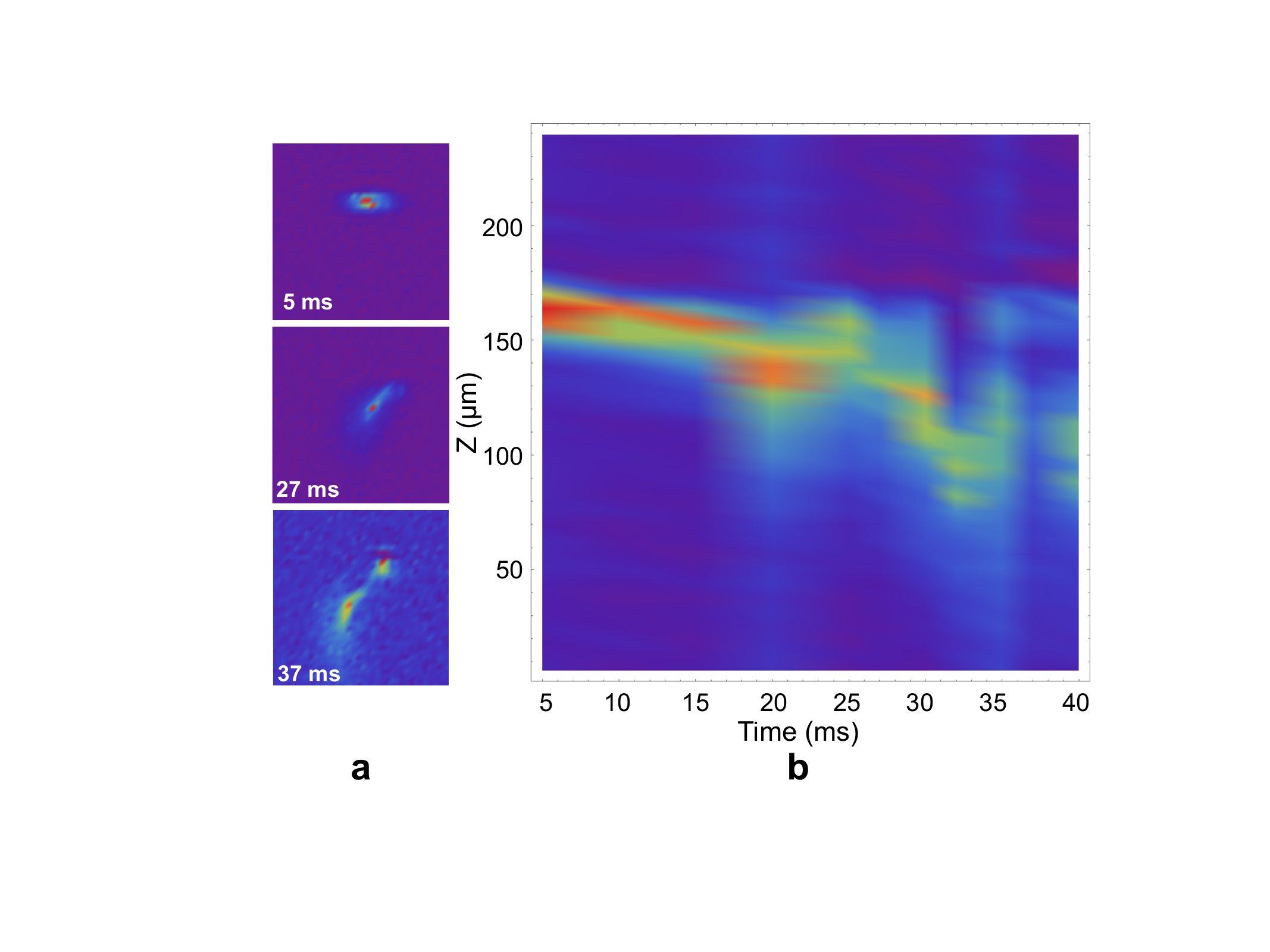}
	\caption{\textbf{Time Evolution of Tunnelling Wave Packet.} \textbf{a} The three stages of the tunnelling process at 5 ms, 27 ms and 37 ms  out-coupling time: the initial stage , tunnelling and final stages respectively. \textbf{b} The sum of the number density along horizontal y-direction presented by false color for all images before and after the tunnelling. The horizontal axis represents out-coupling time of atoms from the reservoir after applying a step-kick. The vertical axis denotes the downward position (along z-direction).}
	\label{fig:sum}
\end{figure}

A series of images was taken with high time resolution from 0 ms to 40 ms, while the first output pulse is completely coming out by exerting a step-kick. In Fig.~\ref{fig:sum}a, the typical images at 5 ms, 27 ms and 37 ms show the BEC distributions for the initial, tunnelling and final stages, respectively. To observe the time evolution of tunnelling process, the images with 2 ms time interval were taken and processed by summing each pixel in y-direction to convert them as an one-dimensional (z-direction) distribution, as shown in Fig. \ref{fig:sum}b. The whole process can be categorised in three stages: the initial stage where the condensate acquires some initial momentum, the tunnelling stage where the tunnelling of atoms occurred, and the final stage where a complete BEC pulse is produced. A steep change in the velocity of the front (lower) end of the tunnelling wave packet appears in the tunnelling stage between 20-30 ms. After the completion of tunnelling, 30-40 ms, the velocity is then slowed down. In contrast to the classical physics, where the potential barrier slows down the tunnelling particle's velocity, the wave packet ``suddenly" emerges from the other side of the barrier, as a wave packet splitting in the quantum tunnelling. Although the tunnelling wave packet grows up within a finite time period, it may look ``instantaneous" and causes the no-tunnelling-time-delay measurement in the attosecond streaking experiment\cite{Sainadh:2019jw}. However, as shown in our experiment, quantum tunnelling has no classical analogue and its entire process can not be properly described by particle transmission\cite{Sokolovski:2018gx}.

\section*{DISCUSSION}
The quantum tunnelling was demonstrated using a hybrid cross optical dipole trap and exerting a non-adiabatic kick, which results in a collective rotational motion in the condensate. Unlike previous work\cite{Potnis:2017dv}, the density dependent chemical potential, the nonlinear term $|\psi|^2$ of the GPE, is negligible under the given energies of the barrier and the kicks. In such cases, the BEC is an ideal platform, not only for many-body quantum physics, but as a giant wave function to study the quantum dynamics of single particles. In our experiment, the detailed process of quantum tunnelling to open space was recorded as a sequence of images in real space. The time evolution of this non-equilibrium dynamics, which was induced by the collective rotational mode of BEC, was then studied. The tunnelling efficiency was found to remain constant, and there is no energy loss of the internal rotational motion during more than 80 ms output time. The tunnelling output pulses remain in BEC phase as indicated by the TOF measurement, and its trace is a temporal map to the internal evolution, as a chronology of its dynamics. The similar methodology has been used to investigate the spatiotemporal motion of valence electron in molecules\cite{Kubel:2019ik, Walt:2017hn}. This result sheds light on the study of tunnelling ionization, which is currently under intensive study using the ultrafast technology, particularly on the ``tunnelling time" problem. As shown in our observation, due to the nature of matter wave in quantum regime, the ``tunnelling time" that can only be well-defined from the point of view of particle, is inadequate\cite{Sokolovski:2018gx}. Instead, the quantum tunnelling process should be described by wavepacket distribution using wave mechanics.

\section*{METHOD}
\subsection{BEC and Trap potential}
Our experimental setup is a hybrid trap\cite{Lin:2009it} together with magnetic transport via a linear motor.  Initially, $^{87}$Rb atoms were loaded to a 3D MOT and then magnetically transported to the UHV science cell through a differential pump of 3 mm diameter. In the science cell, $\sim$ $2.1\times10^8$ atoms were successfully transported with a temperature of 41~$\mu$K. The atomic cloud was then pre-cooled to 8.2~$\mu$K using the RF-knife-cut cooling method. The atoms were subsequently loaded to a hybrid trap that consists of a crossed ODT and a weak magnetic trap to provide additional confinement in the horizontal direction, and partially compensate the gravity. The final stage to reach the quantum degeneracy is the forced-evaporative cooling that was performed  by ramping down the ODT power to a effective trap depth of 580~nK which yielded a Bose-Einstein condensate of $1\times10^5$ atoms with a peak number density 1.02$\times10^{13}$ cm$^{-3}$. The final trapping frequencies were estimated to be: the radial frequencies $\omega_r = 2\pi\times251$ Hz and the axial frequency $\omega_x = 2\pi\times82$ Hz and $\omega_y = 2\pi\times31$ Hz. In our experiment, the location of ODT relative to the center of magnetic quadrupole trap is $(x,y,z)=(0,0,z_0=90~\mu m)$. With the gravity on the z direction, the relevant barrier potential, as shown in Fig.~\ref{fig:timeseq}a, ~\ref{fig:timeseq}b, can be expressed as:
\begin{eqnarray}
U(x,y,z)=-U_t\left\lbrace\exp{\left(-2\frac{x^2 +(z-z_0)^2}{w_0^2}\right)}+\exp{\left(-2\frac{y^2 +(z-z_0)^2}{w_0^2}\right)}\right\rbrace+\beta(x,y,z) 
\label{eq:barrier}
\end{eqnarray}
where the total effective downward force is $\beta(x,y,z)=-m_{Rb}gz+\mu_{Rb}\frac{dB}{dz}\sqrt{\frac{x^2}{4}+\frac{y^2}{4}+z^2}$, $w_0$ is the ODT beam waist, and $U_t$ is the trap depth. The center of atomic cloud corresponding to $(0,0,z_0)$ is then located at the stable point $dU/dz=0$, where $z=z_0-w_0^2\beta/8U_t$ by the approximation of the potential well as a quadratic function. The trapping force in the horizontal X-Y plane given by the magnetic quadrupole provides an extra confinement, while the atoms move out of ODT. The 3D illustration of the potential shape is as Fig.~\ref{fig:timeseq}b.

\subsection{Kick}
After creating BEC in the hybrid trap, a `kick' was then applied to the condensate to perform quantum tunnelling. The `kick' is a sudden change in the ODT power. In our system, two different types of kick were employed. The first type is defined as a `step' kick in which ODT power was reduced to a certain value and maintained during out-coupling time as shown in Fig.~\ref{fig:timeseq}c.  The second type is a `pulse' kick, in which ODT power was increased suddenly for 2 ms and turned back to the previous value.    In order to partially compensate gravity, a magnetic gradient was provided by the compensation quadrupole field to reduce the downward acceleration as low as 0.77 m/s$^2$. The advantage of reduced effective gravitational acceleration is to have a longer observation time by avoiding atom's quick escape out of the field of view. This quadrupole magnetic field also provided a confinement in X-Y plane. 

The sudden change in the ODT power by $\Delta U_t$ is equivalent to shift the stable point (the bottom of the potential well) by $\sim\Delta U_t \frac{w_0^2\beta}{8 U_t^2}$. Because the atomic cloud  cannot follow such quick change, it sits on the original position and gains an energy of $\gamma^2\frac{\beta^2 w_0^2}{8U_t}$, where $\gamma=\Delta U_t/U_t$ is the ODT power change ratio. The kicks shifted the location of potential minima and effectively moves the condensate away from the potential minima.  Then, a collective mode of BEC was excited due to such a non-adiabatic process\cite{Stringari:1996vi, Dalfovo1999}. The step-kick also opens a tunnelling channel on the bottom of the hybrid trap, where the potential barrier is lower because of the gravitational field. Typically, the step-kick is to reduce the trap depth by 30-40\%. 

Additionally, to compare with non-adiabatic kicks, once BEC was produced, then the ODT power was further ramped down adiabatically in two seconds to the trap depth same as the final potential barrier of a typical kick.

\subsection{Internal dynamics TOF measurement}	
A step-kick was applied with a relatively low kick strength, just below the tunnelling threshold, to prevent any observable tunnelling. After a holding time, the BEC was then released from ODT. The position of CM of the atomic cloud was measured with a 10 ms TOF. This was performed for various holding times, from 10~ms to 30~ms with 2~ms time resolution. The periodic oscillation of the CM displacements was found and the variation of the initial velocity after releasing from trap was derived.

\begin{addendum}
\item [Data availability]
The data supporting the results of this study are available from the corresponding author upon reasonable request.
\end{addendum}

\bibliographystyle{naturemag}

\begin{thebibliography}{10}
\expandafter\ifx\csname url\endcsname\relax
  \def\url#1{\texttt{#1}}\fi
\expandafter\ifx\csname urlprefix\endcsname\relax\def\urlprefix{URL }\fi
\providecommand{\bibinfo}[2]{#2}
\providecommand{\eprint}[2][]{\url{#2}}

\bibitem{Holstein:1996il}
\bibinfo{author}{Holstein, B.~R.}
\newblock \bibinfo{title}{{Understanding alpha decay}}.
\newblock \emph{\bibinfo{journal}{American Journal of Physics}}
  \textbf{\bibinfo{volume}{64}}, \bibinfo{pages}{1061--1071}
  (\bibinfo{year}{1996}).

\bibitem{PhysRevD.15.2929}
\bibinfo{author}{Coleman, S.}
\newblock \bibinfo{title}{{fate of the false vacuum: semiclassical theory}}.
\newblock \emph{\bibinfo{journal}{Phys. Rev. D}} \textbf{\bibinfo{volume}{15}},
  \bibinfo{pages}{2929--2936} (\bibinfo{year}{1977}).

\bibitem{JOSEPHSON:1962vk}
\bibinfo{author}{Josephson, B.~D.}
\newblock \bibinfo{title}{{Possible New Effects in Superconductive
  Tunnelling}}.
\newblock \emph{\bibinfo{journal}{Phys Lett}} \textbf{\bibinfo{volume}{1}},
  \bibinfo{pages}{251--253} (\bibinfo{year}{1962}).

\bibitem{AWSCHALOM:1992dl}
\bibinfo{author}{Awschalom, D.~D.}, \bibinfo{author}{Smyth, J.~F.},
  \bibinfo{author}{Grinstein, G.}, \bibinfo{author}{DiVincenzo, D.~P.} \&
  \bibinfo{author}{Loss, D.}
\newblock \bibinfo{title}{Macroscopic quantum tunneling in magnetic proteins}.
\newblock \emph{\bibinfo{journal}{Phys. Rev. Lett.}}
  \textbf{\bibinfo{volume}{68}}, \bibinfo{pages}{3092--3095}
  (\bibinfo{year}{1992}).

\bibitem{Collini:2010fy}
\bibinfo{author}{Collini, E.} \emph{et~al.}
\newblock \bibinfo{title}{{Coherently wired light-harvesting in photosynthetic
  marine algae at ambient temperature}}.
\newblock \emph{\bibinfo{journal}{Nature}} \textbf{\bibinfo{volume}{463}},
  \bibinfo{pages}{1--5} (\bibinfo{year}{2010}).

\bibitem{Eckle:2008eu}
\bibinfo{author}{Eckle, P.} \emph{et~al.}
\newblock \bibinfo{title}{{Attosecond angular streaking}}.
\newblock \emph{\bibinfo{journal}{Nat Phys}} \textbf{\bibinfo{volume}{4}},
  \bibinfo{pages}{565--570} (\bibinfo{year}{2008}).

\bibitem{Pfeiffer:2011kl}
\bibinfo{author}{Pfeiffer, A.~N.} \emph{et~al.}
\newblock \bibinfo{title}{{Attoclock reveals natural coordinates of the
  laser-induced tunnelling current flow in atoms}}.
\newblock \emph{\bibinfo{journal}{Nat Phys}} \textbf{\bibinfo{volume}{8}},
  \bibinfo{pages}{1--5} (\bibinfo{year}{2011}).

\bibitem{Walt:2017hn}
\bibinfo{author}{Walt, S.~G.} \emph{et~al.}
\newblock \bibinfo{title}{{Dynamics of valence-shell electrons and nuclei
  probed by strong-field holography and rescattering.}}
\newblock \emph{\bibinfo{journal}{Nat Comms}} \textbf{\bibinfo{volume}{8}},
  \bibinfo{pages}{15651} (\bibinfo{year}{2017}).

\bibitem{Kubel:2019ik}
\bibinfo{author}{K{\"u}bel, M.} \emph{et~al.}
\newblock \bibinfo{title}{{Spatiotemporal imaging of valence electron motion}}.
\newblock \emph{\bibinfo{journal}{Nat Comms}} \textbf{\bibinfo{volume}{10}},
  \bibinfo{pages}{1--7} (\bibinfo{year}{2019}).

\bibitem{Sainadh:2019jw}
\bibinfo{author}{Sainadh, U.~S.} \emph{et~al.}
\newblock \bibinfo{title}{{Attosecond angular streaking and tunnelling time in
  atomic hydrogen}}.
\newblock \emph{\bibinfo{journal}{Nature}} \textbf{\bibinfo{volume}{568}},
  \bibinfo{pages}{1--9} (\bibinfo{year}{2019}).

\bibitem{Wu:2019fu}
\bibinfo{author}{Wu, J.} \emph{et~al.}
\newblock \bibinfo{title}{{Probing the tunnelling site of electrons in strong
  field enhanced ionization of molecules.}}
\newblock \emph{\bibinfo{journal}{Nat Comms}} \textbf{\bibinfo{volume}{3}},
  \bibinfo{pages}{1113} (\bibinfo{year}{2012}).

\bibitem{Torlina:2015dl}
\bibinfo{author}{Torlina, L.} \emph{et~al.}
\newblock \bibinfo{title}{{Interpreting attoclock measurements of tunnelling
  times}}.
\newblock \emph{\bibinfo{journal}{Nat Phys}} \textbf{\bibinfo{volume}{11}},
  \bibinfo{pages}{503--508} (\bibinfo{year}{2015}).

\bibitem{Ni:2016hq}
\bibinfo{author}{Ni, H.}, \bibinfo{author}{Saalmann, U.} \&
  \bibinfo{author}{Rost, J.-M.}
\newblock \bibinfo{title}{Tunneling ionization time resolved by
  backpropagation}.
\newblock \emph{\bibinfo{journal}{Phys. Rev. Lett.}}
  \textbf{\bibinfo{volume}{117}}, \bibinfo{pages}{023002}
  (\bibinfo{year}{2016}).

\bibitem{Sokolovski:2018gx}
\bibinfo{author}{Sokolovski, D.} \& \bibinfo{author}{Akhmatskaya, E.}
\newblock \bibinfo{title}{{No time at the end of the tunnel}}.
\newblock \emph{\bibinfo{journal}{Communications Physics}}
  \textbf{\bibinfo{volume}{1}}, \bibinfo{pages}{1--9} (\bibinfo{year}{2018}).

\bibitem{Anderson:1998uf}
\bibinfo{author}{Anderson, B.~P.} \& \bibinfo{author}{Kasevich, M.~A.}
\newblock \bibinfo{title}{{Macroscopic quantum interference from atomic tunnel
  arrays}}.
\newblock \emph{\bibinfo{journal}{Science}} \textbf{\bibinfo{volume}{282}},
  \bibinfo{pages}{1686--1689} (\bibinfo{year}{1998}).

\bibitem{Serwane:2011cv}
\bibinfo{author}{Serwane, F.} \emph{et~al.}
\newblock \bibinfo{title}{{Deterministic preparation of a tunable few-fermion
  system.}}
\newblock \emph{\bibinfo{journal}{Science}} \textbf{\bibinfo{volume}{332}},
  \bibinfo{pages}{336--338} (\bibinfo{year}{2011}).

\bibitem{Lode:2012kr}
\bibinfo{author}{Lode, A. U.~J.}, \bibinfo{author}{Streltsov, A.~I.},
  \bibinfo{author}{Sakmann, K.}, \bibinfo{author}{Alon, O.~E.} \&
  \bibinfo{author}{Cederbaum, L.~S.}
\newblock \bibinfo{title}{{How an interacting many-body system tunnels through
  a potential barrier to open space}}.
\newblock \emph{\bibinfo{journal}{Proceedings of the National Academy of
  Sciences of the United States of America}} \textbf{\bibinfo{volume}{109}},
  \bibinfo{pages}{13521--13525} (\bibinfo{year}{2012}).

\bibitem{Lignier2007}
\bibinfo{author}{Lignier, H.} \emph{et~al.}
\newblock \bibinfo{title}{Dynamical control of matter-wave tunneling in
  periodic potentials}.
\newblock \emph{\bibinfo{journal}{Phys. Rev. Lett.}}
  \textbf{\bibinfo{volume}{99}}, \bibinfo{pages}{220403}
  (\bibinfo{year}{2007}).

\bibitem{Tayebirad:2010}
\bibinfo{author}{Tayebirad, G.} \emph{et~al.}
\newblock \bibinfo{title}{Time-resolved measurement of landau-zener tunneling
  in different bases}.
\newblock \emph{\bibinfo{journal}{Phys. Rev. A}} \textbf{\bibinfo{volume}{82}},
  \bibinfo{pages}{013633} (\bibinfo{year}{2010}).

\bibitem{Bloch:2008ch}
\bibinfo{author}{Bloch, I.}, \bibinfo{author}{Dalibard, J.} \&
  \bibinfo{author}{Zwerger, W.}
\newblock \bibinfo{title}{{Many-body physics with ultracold gases}}.
\newblock \emph{\bibinfo{journal}{Rev Mod Phys}} \textbf{\bibinfo{volume}{80}},
  \bibinfo{pages}{885--964} (\bibinfo{year}{2008}).

\bibitem{Potnis:2017dv}
\bibinfo{author}{Potnis, S.}, \bibinfo{author}{Ramos, R.},
  \bibinfo{author}{Maeda, K.}, \bibinfo{author}{Carr, L.~D.} \&
  \bibinfo{author}{Steinberg, A.~M.}
\newblock \bibinfo{title}{Interaction-assisted quantum tunneling of a
  bose-einstein condensate out of a single trapping well}.
\newblock \emph{\bibinfo{journal}{Phys. Rev. Lett.}}
  \textbf{\bibinfo{volume}{118}}, \bibinfo{pages}{060402}
  (\bibinfo{year}{2017}).

\bibitem{Alcala:2017jj}
\bibinfo{author}{Alcala, D.~A.}, \bibinfo{author}{Glick, J.~A.} \&
  \bibinfo{author}{Carr, L.~D.}
\newblock \bibinfo{title}{{Entangled Dynamics in Macroscopic Quantum Tunneling
  of Bose-Einstein Condensates}}.
\newblock \emph{\bibinfo{journal}{Phys. Rev. Lett.}}
  \textbf{\bibinfo{volume}{118}}, \bibinfo{pages}{210403}
  (\bibinfo{year}{2017}).

\bibitem{Barak:2008bk}
\bibinfo{author}{Barak, A.}, \bibinfo{author}{Peleg, O.},
  \bibinfo{author}{Stucchio, C.}, \bibinfo{author}{Soffer, A.} \&
  \bibinfo{author}{Segev, M.}
\newblock \bibinfo{title}{Observation of soliton tunneling phenomena and
  soliton ejection}.
\newblock \emph{\bibinfo{journal}{Phys. Rev. Lett.}}
  \textbf{\bibinfo{volume}{100}}, \bibinfo{pages}{153901}
  (\bibinfo{year}{2008}).

\bibitem{Salasnich:2002dr}
\bibinfo{author}{Salasnich, L.}, \bibinfo{author}{Parola, A.} \&
  \bibinfo{author}{Reatto, L.}
\newblock \bibinfo{title}{{Periodic quantum tunnelling and parametric resonance
  with cigar-shaped Bose-Einstein condensates}}.
\newblock \emph{\bibinfo{journal}{J Phys B-At Mol Opt}}
  \textbf{\bibinfo{volume}{35}}, \bibinfo{pages}{3205--3216}
  (\bibinfo{year}{2002}).

\bibitem{griffiths:quantum}
\bibinfo{author}{Griffiths, D.}
\newblock \emph{\bibinfo{title}{Introduction of Quantum Mechanics}}
  (\bibinfo{publisher}{Prentice Hall, Inc.}, \bibinfo{year}{1995}).

\bibitem{Chen:2005hl}
\bibinfo{author}{Chen, P. Y.~P.} \& \bibinfo{author}{Malomed, B.~A.}
\newblock \bibinfo{title}{{A model of a dual-core matter-wave soliton laser}}.
\newblock \emph{\bibinfo{journal}{J Phys B-At Mol Opt}}
  \textbf{\bibinfo{volume}{38}}, \bibinfo{pages}{4221--4234}
  (\bibinfo{year}{2005}).

\bibitem{Stringari:1996vi}
\bibinfo{author}{Stringari, S.}
\newblock \bibinfo{title}{{Collective excitations of a trapped bose-condensed
  gas}}.
\newblock \emph{\bibinfo{journal}{Phys. Rev. Lett.}}
  \textbf{\bibinfo{volume}{77}}, \bibinfo{pages}{2360--2363}
  (\bibinfo{year}{1996}).

\bibitem{Dalfovo1999}
\bibinfo{author}{Dalfovo, F.}, \bibinfo{author}{Giorgini, S.},
  \bibinfo{author}{Pitaevskii, L.~P.} \& \bibinfo{author}{Stringari, S.}
\newblock \bibinfo{title}{Theory of bose-einstein condensation in trapped
  gases}.
\newblock \emph{\bibinfo{journal}{Rev. Mod. Phys.}}
  \textbf{\bibinfo{volume}{71}}, \bibinfo{pages}{463--512}
  (\bibinfo{year}{1999}).

\bibitem{HAUGE:1989tm}
\bibinfo{author}{Hauge, E.~H.} \& \bibinfo{author}{Stovneng, J.~A.}
\newblock \bibinfo{title}{{Tunneling Times - a Critical-Review}}.
\newblock \emph{\bibinfo{journal}{Rev Mod Phys}} \textbf{\bibinfo{volume}{61}},
  \bibinfo{pages}{917--936} (\bibinfo{year}{1989}).

\bibitem{Lin:2009it}
\bibinfo{author}{Lin, Y.-J.}, \bibinfo{author}{Perry, A.~R.},
  \bibinfo{author}{Compton, R.~L.}, \bibinfo{author}{Spielman, I.~B.} \&
  \bibinfo{author}{Porto, J.~V.}
\newblock \bibinfo{title}{Rapid production of $^{87}\text{R}\text{b}$
  bose-einstein condensates in a combined magnetic and optical potential}.
\newblock \emph{\bibinfo{journal}{Phys. Rev. A}} \textbf{\bibinfo{volume}{79}},
  \bibinfo{pages}{063631} (\bibinfo{year}{2009}).

\end{thebibliography}

\begin{addendum}
\item [Acknowledgements]
This work was financially supported by the Center for Quantum Technology from the Featured Areas Research Center Program within the framework of the Higher Education Sprout Project by the Ministry of Education (MOE) and the Ministry of science and Technology (MOST), Taiwan (Grant No. 106-2112-M-007-021-MY3 and 105-2112-M-007-027-MY3)

\item [Author Contributions]
Y.W.L. and K.S. conceived the experiment. J.R.C. construct the experimental apparatus. K.S. and P.S.C. performed the measurements. J.R.C. and Y.H.C. helped in taking data.  Y.W.L. and K.S. analyzed the data.  Y.W.L. and K.S. wrote the manuscript. All the authors contributed to the experiment.

\item [Competing interests]
The authors declare no competing interests.

\end{addendum}

\end{document}